\documentclass[nofootinbib,letterpaper,preprint,preprintnumbers,amsmath,amssymb]{revtex4}
\usepackage{latexsym}
\usepackage{hyperref}
\begin{document}
\title{\bf Parallel Symbolic Computation of Curvature Invariants in General Relativity}
\author{K. R. Koehler}
\affiliation{University of Cincinnati / Raymond Walters College, Cincinnati, OH 45236}
\email{kenneth.koehler@uc.edu}
\begin{abstract}
We present a practical application of parallel symbolic computation in General Relativity:
the calculation of curvature invariants for large dimension. We discuss the structure
of the calculations, an implementation of the technique and scaling of the computation 
with spacetime dimension for various invariants.
\end{abstract} \maketitle
Parallel symbolic computation has received increasing attention since the early 1990's, and a
number of applications in pure mathematics \cite{Siegl} \cite{Hitz} and databases \cite{Waltz} have been explored.
Little attention, however, has been paid to possible applications in General Relativity. This note
describes such an application.

Curvature invariants are scalar products of Riemann, Ricci or Weyl Tensors or their covariant derivatives.
The canonical example is the Kretschmann Invariant
\begin{eqnarray}
R^{a b c d} R_{a b c d}
\end{eqnarray}
Here Einstein Summation (also known as index contraction) is performed on all repeated indices. 
The Kretschmann Invariant is the simplest invariant product involving the Riemann Curvature Tensor.  
It is used most often to identify essential singularities in a spacetime geometry.

A variety of other curvature invariants have been considered in the literature. For example
\begin{eqnarray}
I_1 = R^{a b c d; e f} R_{a g c h; e f} R^{i g j h; k l} R_{i b j d; k l}
\end{eqnarray}
(where the semi-colon denotes covariant differentiation) 
has been used \cite{Bicak} to analyze type N spacetimes. The same authors \cite{Bicak2} used
\begin{eqnarray}
I_2 = R^{a b c d} R_{a e f g} R^{e f}\ _{b h} R^{g h}\ _{c d}
\end{eqnarray}
to locate asymptotically flat regions of the spinning C metric.
These calculations all share two common characteristics:
\begin{itemize}
\item they are sums of products of tensor components whose indices can all be enumerated before the 
calculation begins, and 
\item the final results are relatively simple sums of multinomials. 
\end{itemize}
For example, $I_2$ for the spinning C metric is \cite{Bicak2}
\begin{eqnarray}
- 144 m^4 \frac{(p + q)^{12}}{(1 + p^2 q^2)^6}.
\end{eqnarray}

Since the set of indices occurring in the sum is enumerable before computation begins, the sum
can easily be partitioned and carried out on multiple processors. Because the metric is typically a function
of a small number of variables, the number of different multinomials
occurring in the sum is correspondingly small relative to the number of products which must be summed.
This means that partial sums can be accumulated on each processor, and when all of the partial sums
are complete, final summation is in general a short process. For the same reason, simplification
of the final sum is relatively straightforward.

This application of the Divide and Conquer Principle
\cite{Roosta} lends itself well to loosely coupled multiprocessor configurations such as the Beowulf Cluster 
\cite{Reschke} \cite{Ridge} because the close similarities between the partial sums lead to an essentially linear
scaling of execution time with cluster size. It is important to note that as the number of processors increases, 
the sizes of the partial sums may increase as terms are accumulated in different processors which will cancel 
in the final sum. This increase is in general small, but is dependent on the specific metric and
invariant studied. Since the sizes of the various partial sums are not known a prioi, a slight overall
decrease in execution time can be achieved by partitioning the workload into many more parcels than
processors.

In $D$ dimensions, each summation index can take $D$ values, so that an invariant with $n$ summed
indices requires the summation of at most $D^n$ products. This number can often be reduced by consideration
of the symmetries of the Riemann Curvature Tensor:
\begin{eqnarray}
R_{a b c d} &=& - R_{b a c d} \cr
&=& - R_{a b d c} \cr
&=& R_{c d a b}
\end{eqnarray}
The first two of these symmetries implies that pairs of antisymmetric indices which are summed together 
(for instance, $c$ and $d$ in $I_2$) contribute a factor of $\frac{D (D - 1)}{2}$ to the number of products.
The last symmetry is only useful for specific invariants such
as the Kretschmann Invariant, where it reduces the number of products by the factor
\begin{eqnarray}
\frac{2 D (D - 1)}{D (D - 1) + 2}
\end{eqnarray}
Summation over pairs of symmetric indices (such as $e$ and $f$ or $k$ and $l$ in $I_1$) contribute
a factor of $\frac{D (D + 1)}{2}$ to the number of products. Sums over lone indices contribute a factor of $D$.
It must also be remembered that for any given metric, \emph{not all of the Riemann Tensor components are
nonzero}. Hence the computation of $I_1$ involves the sum of \emph{at most}
\begin{eqnarray}
\frac{D^{10} (D + 1)^2}{4}
\end{eqnarray}
products quartic in second derivatives of the Riemann Tensor, each of which may contain tens of terms.
In four dimensions, this is 6,553,600 products and perhaps tens or hundreds of billions of terms,
depending on the complexity of the metric. In comparison, the worst case scenario for the computation 
of $I_2$ requires the sum of
\begin{eqnarray}
\frac{D^7 (D - 1)}{2}
\end{eqnarray}
products quartic in the Riemann Tensor.

The algorithm for index enumeration is straightforward if a bit messy.
Assume that $tensor\_array$ is an array of tensors occuring in the invariant and that
all necessary index raising has already been accomplished. For instance, in $I_2$, the elements of
$tensor\_array$ would be
\begin{quote}
$R^{a b c d}, R_{a b c d}, R^{a b}\ _{c d}$ and $R^{a b}\ _{c d}$.
\end{quote}
The tensor data structure must contain, in addition to a sparse array containing the
nonzero components for that tensor, an array indicating which pairs of indices are
antisymmetric, and an array $indices$ which indicates which indices are summed together.
For $I_2$, $indices$ for each tensor would be
\begin{quote}
$\{1, 2, 3, 4\}, \{1, 5, 6, 7\}, \{5, 6, 2, 8\}$ and $\{7, 8, 3, 4\}$,
\end{quote}
indicating that, for instance, the second index of the first tensor should be summed over with the
third index of the third tensor, etc.

The following pseudocode details the enumeration of the indices of the component products to be summed
with respect to antisymmetric index pairs as well as the elimination of products involving zero
components. We begin by figuring out which pairs of sums can be abbreviated due to symmetry
considerations:
\begin{quote}
\begin{tabbing}
$multiplier$ = 1\\
for each pair of tensors $A$ and $B$ \{ \\
\quad \=   for each pair of $indices (I, I+1)$ in $A$ and $(J,J+1)$ in $B$ \{ \\
\> \quad \=      if $A.indices(I) == B.indices(J)$ and $A.indices(I+1) == B.indices(J+1)$ \{ \\
\> \> \quad \=         if both pairs are antisymmetric \{ \\
\> \> \> \quad \=            remember to abbreviate summation on this pair of indices\\
\> \> \> \>                  $multiplier$ = $multiplier$ * 2 \} \} \} \}
\end{tabbing}
\end{quote}
Summation abbreviation is accomplished by ignoring index pairs in which the second index of
each pair is less than the first. Each product will be multiplied by $multiplier$ to correct
for the abbreviation. We chose not to implement summation abbreviation for
symmetric pairs of indices, or for larger sets of antisymmetric indices, for reasons of
simplicity. We also chose to ignore the third symmetry property of the Riemann Tensor, because it applies
relatively rarely in the types of invariants which might benefit from parallel processing.

We can now enumerate the indices which will contribute to the invariant. The array $sum\_indices$ 
contains the set of indices under consideration for a given product. We assume
that the range of indices is from $0$ to $D - 1$, and that as a given index increments to $D$,
it is set to zero and the next index is incremented by 1. In the following we will refer to this procedure
as ``cycling the indices''. The resulting $sum\_index\_array$ is the complete list
of all products which contribute to the invariant:
\begin{quote}
\begin{tabbing}
set $sum\_indices$ to all zeroes\\
$product\_count$ = 0\\
do \{ \\
\quad \= $process\_these\_indices = TRUE$\\
\>       for each index pair $\{I, J\}$ \{ \\
\> \quad \=   if abbreviating summation on this pair \{ \\
\> \> \quad \=     $process\_these\_indices = process\_these\_indices$ AND $(J > I)$ \} \} \\
\>       if $process\_these\_indices$ \{ \\
\> \quad \=   if all tensor component corresponding to these indices are nonzero \{ \\
\> \> \quad \=     $sum\_index\_array(product\_count) = sum\_indices$\\
\> \> \>           $product\_count = product\_count + 1$ \} \} \\
\>   cycle $sum\_indices$ \} \\
until $sum\_indices$ cycles back to all zeroes
\end{tabbing}
\end{quote}

The parallelization of curvature invariant computation can then be summarized as follows, assuming
$n$ processors with $m$ parcels per processor:
\begin{enumerate}
\item compute the components of the Riemann Tensor and any derivatives which occur in the invariant;
\item enumerate the tensor indices occurring in the sum as described above;
\item partition $sum\_index\_array$ into $m * n$ parcels;
\item distribute new parcels to processors as they become idle, with each processor accumulating a partial sum;
\item retrieve the $n$ partial sums and simplify the full sum.
\end{enumerate}
This algorithm has been implemented by the author in a prototype code set called PTAH 
(Parallel Tensor Algebra Hybrid) \cite{Koehler}, which is a hybrid of C++ and Mathematica.
The results below were obtained using that code, and were then cross-checked using Mathematica
alone on a single processor. The enumeration pseudocode above is derived from the corresponding code in PTAH.

To examine the scaling properties of this technique, we require a related class of spacetime metrics
in diverse dimensions. One such class is that of the Kerr Metrics with single rotation parameter \cite{Myers}.
Let
\begin{eqnarray}
\rho^2 &=& r^2 + a^2 \cos^2 \theta
\cr
\Delta_D &=& \frac{\mu}{r^{D-5} \rho^2}
\cr
\Psi_D &=& \frac{\rho^2}{(r^2 + a^2) - \frac{\mu}{r^{D-5}}}
\end{eqnarray}
where $\mu$ and $a$ are proportional to the mass and angular momentum respectively.
Then the metric in $D$ dimensions is
\begin{eqnarray}
ds^2 &=& r^2 \cos^2 \theta d\Omega^2 + \Psi_D dr^2 + \rho^2 d\theta^2 + 
\cr
&&((r^2 + a^2) \sin^2 \theta + \Delta_D a^2 \sin^4 \theta) d\phi^2 + 
\cr
&&2 \Delta_D a \sin^2 \theta d\phi dt + (\Delta_D - 1) dt^2
\end{eqnarray}
where $d\Omega^2$ is the standard metric on $S^{D-4}$. We note that the nonzero Riemann Tensor components
for this metric in $D$ = 4 have between 2 and 18 terms. In $D$ = 6 and 8,
the Riemann Tensor components have as many as 25 terms. For a general metric in $D$ dimensions, there are
\begin{eqnarray}
\frac{D^2 (D^2 - 1)}{12} 
\end{eqnarray}
independent curvature components. For the Kerr metric in $D$ = 4, 13 of the possible 21
independent components are nonzero. In contrast, for $D$ = 11, the Kerr curvature has only 68 nonzero
independent components of the possible 1210.

The Kerr Metrics depend explicitly on $D-1$ variables: $r, \theta, a, \mu$ and the $D-5$ ``polar'' angles
parametrizing the $S^{D-4}$. We summarize the computations of three invariants:
\begin{eqnarray}
I_a &=& R^{a b c d} R_{a b c d} \cr
I_b &=& R^{a b c d} R^{e f}\ _{a b} R_{c d e f} \cr
I_c &=& R^{a b c d; e} R_{a b c d; e}
\end{eqnarray}
for $D=4$ to $D=11$ in the following table \cite{Koehler2}:
\begin{center}
\begin{tabular}[c]{r r r r r r r r r r r r r r r r r r r}
$D$ & & & $P(I_a)$ & & & $P(I_b)$ & & & $P(I_c)$ & & & $T(I_a)$ & & & $T(I_b)$ & & & $T(I_c)$ \cr
 4  & & &    172   & & &    244   & & &   1160   & & &      4   & & &      5   & & &     11 \cr 
 5  & & &    224   & & &    309   & & &   1919   & & &      3   & & &      4   & & &     12 \cr 
 6  & & &    290   & & &    391   & & &   2749   & & &      5   & & &      6   & & &     21 \cr 
 7  & & &    373   & & &    495   & & &   3982   & & &      5   & & &      7   & & &     21 \cr 
 8  & & &    477   & & &    628   & & &   5703   & & &      5   & & &      7   & & &     21 \cr 
 9  & & &    606   & & &    798   & & &   8109   & & &      5   & & &      7   & & &     21 \cr 
10  & & &    764   & & &   1014   & & &  11390   & & &      5   & & &      7   & & &     21 \cr 
11  & & &    955   & & &   1286   & & &  15769   & & &      5   & & &      7   & & &     21 \cr 
\end{tabular}
\end{center}
Here $P(I)$ is the number of products of tensor components in the sum and $T(I)$ is the number
of terms in the final result. The $P(I)$ includes those operations necessary to raise
tensor components before index contraction is possible, but excludes all unnecessary
products due to zero components and symmetry considerations.

Even with the rise in the number of symbolic multiplications with spacetime
dimension and the substantial variation between different invariants, we see that for this class
of examples the final sums are quite small. This indicates that this technique can be efficiently
parallelized under a loosely coupled MIMD architecture. We have also investigated a set of supergravity brane
metrics in 10 and 11 dimensions \cite{Koehler3} which produced even smaller final sums 
(in accordance with the relatively simpler metric structures involved).

While they illustrate the salient features of the computational technique, the Kerr invariants discussed 
here are not sufficiently complicated to warrant parallelization on a modern Beowulf Cluster.
A more challenging computation is the Euler Class, which in 10 dimensions is quintic in the
Riemann Tensor and requires (worst case) the sum of over 6 billion products. For the Kerr metric,
the result has but 4 terms. For sufficiently complicated metrics and large $D$, 
invariants like $I_1$ and $I_2$ may not
be practically computable in any other way but in parallel, but because of the small number of variables
parametrizing the metric, the results are expected to be similar to those described here.

It is also possible to use this technique to perform index contractions in tensor products which are
\emph{not} scalars. In this case the contracted indices are enumerated once for each set of values of
the free (uncontracted) indices. The technique is even more effective in spinor contractions. Since the
number of spinor components in $D$ dimensions is $2^{Floor D / 2}$, the number of products occurring
in each sum grows much more quickly with dimension than in the tensor case.

\section*{Acknowledgments}
We would like to thank Paul Esposito, Richard Gass, Mark Jarrell, Cenalo Vaz and Louis Witten for 
stimulating discussions on these matters.

\end{document}